\documentclass[a4paper,11pt]{article}
\pdfoutput=1
\usepackage{jheppub}
\usepackage{amsmath,amssymb}
\usepackage[usenames,dvipsnames]{xcolor}
\usepackage{tikz}
\usetikzlibrary{patterns}
\usetikzlibrary{decorations.pathmorphing}
\usetikzlibrary{decorations.markings}
\usepgflibrary{arrows.meta}
\tikzset{
  WLBE/.style={double distance=1.1pt,postaction={decorate}, decoration={markings,mark=at position .1 with {\arrow{Straight Barb[scale=0.5]}},mark=at position .98 with {\arrow{Straight Barb[scale=0.5]}}}},
  WLB/.style={double distance=1.1pt,postaction={decorate},decoration={markings,mark=at position .2 with {\arrow{Straight Barb[scale=0.5]}}}},
  WLBS/.style={double distance=1.1pt,postaction={decorate},decoration={markings,mark=at position .1 with {\arrow{Straight Barb[scale=0.5]}}}},
  GLUON/.style={decorate,decoration={coil, amplitude=2.5pt,segment length=3.25pt, aspect=0.65}}
}
\definecolor{colA}{HTML}{c19277}
\definecolor{colB}{HTML}{e1bc91}
\definecolor{colD}{HTML}{62959c}
\hypersetup{
  citecolor=colA, 		
  linkcolor=colA,    	
  urlcolor=colD,			
}

\newcommand{\ep}{\ensuremath{\varepsilon}}
\newcommand{\MS}{\ensuremath{\overline{\mathrm{MS}}}}
\newcommand{\Zhh}{\ensuremath{Z_{\mathrm{hh}}}}
\newcommand{\Zghh}{\ensuremath{Z_{\mathrm{ghh}}}}
\newcommand{\Zhhhh}{\ensuremath{Z_{\mathrm{[hh]hh}}}}
\newcommand{\Ghh}{\ensuremath{G_{\mathrm{hh}}}}
\newcommand{\Gghh}{\ensuremath{G_{\mathrm{ghh}}}}
\newcommand{\Ghhhh}{\ensuremath{G_{\mathrm{[hh]hh}}}}

\newcommand{\aS}{\ensuremath{a_s}}
\newcommand{\aXi}{\ensuremath{a_\xi}}

\newcommand{\cusp}{\ensuremath{\Gamma_{\mathrm{cusp}}}}
\newcommand{\Zcusp}{\ensuremath{Z_{\mathrm{cusp}}}}

\colorlet{colfact}{MidnightBlue}
\newcommand{\CF}{\ensuremath{\textcolor{colfact}{C_F}}}
\newcommand{\CFCF}{\ensuremath{\textcolor{colfact}{C_F^2}}}
\newcommand{\CFCFCF}{\ensuremath{\textcolor{colfact}{C_F^3}}}
\newcommand{\CA}{\ensuremath{\textcolor{colfact}{C_A}}}
\newcommand{\CACA}{\ensuremath{\textcolor{colfact}{C_A^2}}}
\newcommand{\CACACA}{\ensuremath{\textcolor{colfact}{C_A^3}}}
\newcommand{\TFNF}{\ensuremath{\textcolor{colfact}{n_f T_F}}}
\newcommand{\NF}{\ensuremath{\textcolor{colfact}{n_f}}}
\newcommand{\NC}{\ensuremath{\textcolor{colfact}{N}}}
\newcommand{\dFdF}{\ensuremath{\textcolor{colfact}{d^{abcd}_Fd^{abcd}_F}}}
\newcommand{\dFdA}{\ensuremath{\textcolor{colfact}{d^{abcd}_Fd^{abcd}_A}}}
\newcommand{\BSYM}{\ensuremath{B^{\mathcal{N}=4}}}

\title{Four-loop QCD cusp anomalous dimension at small angle}
\author[a]{Andrey G.~Grozin}
\author[a]{Roman N.~Lee}
\author[b]{and Andrey F.~Pikelner}
\affiliation[a]{Budker Institute of Nuclear Physics,
  Lavrentiev st.~11, Novosibirsk 630090, Russia}
\affiliation[b]{Joint Institute of Nuclear Research,
  Joliot-Curie st.~6, Dubna 141980, Russia}
\emailAdd{A.G.Grozin@inp.nsk.su}
\emailAdd{R.N.Lee@inp.nsk.su}
\emailAdd{pikelner@theor.jinr.ru}
\abstract{We calculate the small angle expansion of the four-loop QCD cusp
  anomalous dimension. As a byproduct of our calculation, we also obtain the
  four-loop anomalous dimension of the heavy-quark field in HQET. The validity of
  the calculational setup is cross-checked by the independent calculation of the
  four-loop QCD beta-function from heavy-quark-gluon vertex renormalization in
  HQET. We check the obtained results for the cusp anomalous dimension and
  heavy-quark field anomalous dimension against available analytical and numerical
  results. Finally, we find that the maximal transcendentality contribution to the
  QCD Bremsstrahlung function coincides, up to a factor $3/2$, with the
  Bremsstrahlung function in $\mathcal{N}=4$ supersymmetric Yang-Mills theory, at
  least, through 4 loops.}
\keywords{Wilson, 't Hooft and Polyakov loops,
  Higher-Order Perturbative Calculations,
  Effective Field Theories of QCD}
\begin{document}
\maketitle
\flushbottom

\section{Introduction}
\label{sec:Intro}
The cusp anomalous dimension $\cusp$ was originally introduced in Polyakov's
paper \cite{Polyakov:1980ca} to describe the scaling of a cusped Wilson loop
with a variation of UV cutoff parameter. The same quantity determines
the infrared singularity structure of scattering amplitudes in the chosen QFT
theory. The physical meaning of this correspondence is simple. Since the
infrared divergencies stem from soft regions of loop integration, the
incoming/outgoing particle can be replaced by a cusped Wilson line which IR and UV
scaling behavior is governed by the same exponent by dimensional arguments. The
notation $\cusp(\phi,\alpha_s)$ indicates that this quantity depends on cusp
angle $\phi$ and coupling constant $\alpha_s$. Of course, the cusp
anomalous dimension depends also on the specific variant of quantum field
theory. In the case of quantum electrodynamics without massless fermions, the
cusp anomalous dimension determines the scaling of quasi-elastic cross sections
with the soft-photon energy cut-off thus being a directly observable quantity.
Note that in this case result exact in $\alpha$ can be obtained from
one-loop calculation, thanks to exponentiation of QED. For non-abelian theories,
in particular, for QCD, $\cusp(\phi,\alpha_s)$ is a nontrivial series in
$\alpha_s$ accessible only via perturbative calculations.
The asymptotics of $\cusp(\phi,\alpha_s)$ at large and small angles also provide
an important information. The light-like cusp anomalous dimension
$K(\alpha_s)=\lim_{\phi\to i \infty} i \cusp(\phi,\alpha_s)/\phi$ plays
important role for the infrared asymptotics of massless scattering amplitudes
and form factors~\cite{Korchemsky:1985xj,Collins:1989gx,Collins:1989bt}. The
opposite limit of small angles is also of some interest. In particular, the
Bremsstrahlung function $B(\alpha_s)=-\lim_{\phi\to 0}
\cusp(\phi,\alpha_s)/\phi^2$ determines the energy loss of a charged particle
moving along a smooth curved trajectory \cite{Correa:2012at}. In addition, the
calculations in the small-angle limit are much more accessible and often precede
the full-angle dependence calculations.
The exact angle dependence of $\cusp(\phi,\alpha_s)$ is known up to three loops in
QCD~\cite{Grozin:2014hna} and supersymmetric Yang-Mills
theories~\cite{Grozin:2015kna}. At the four loop order, only partial results are
available: fermionic contributions to QCD $\cusp(\phi,\alpha_s)$ in small-angle
expansion~\cite{Grozin:2016ydd,Grozin:2017css,Grozin:2018vdn,Bruser:2019auj},
abelian part with full angle
dependence~\cite{Grozin:2016ydd,Grozin:2018vdn,Bruser:2020bsh}, planar part of
angle-dependent cusp anomalous dimension in N=4 SYM~\cite{Henn:2013wfa}. In
addition, the Bremsstrahlung function has been calculated at the four-loop level
in 3-dimensional ABJM theory, Ref.~\cite{Bianchi:2017afp,Bianchi:2017ujp}.
The goal of the present paper is to provide the small-angle expansion up to
$\phi^4$ of the QCD four-loop cusp anomalous dimension. As a byproduct, we
calculate the anomalous dimension of the heavy quark field in heavy quark
effective theory (HQET) \cite{Neubert:1993mb}, extending the partial results of
Refs.~\cite{Grozin:2016ydd,Grozin:2017css,Marquard:2018rwx,Grozin:2018vdn,Bruser:2019auj}.
Both calculations require the knowledge of the four-loop HQET propagator-type
integrals and provide the first application examples of the results of
Ref.~\cite{Lee:2022} where a full set of the four-loop HQET propagator master
integrals was calculated. As an additional cross-check for all ingredients of
the calculation chain we obtain from the renormalization of the
heavy-quark-gluon vertex the four-loop QCD beta-function known for a long
time~\cite{vanRitbergen:1997va,Czakon:2004bu}.
The paper is organized as follows. In section~\ref{sec:HQET-setup} we introduce
the HQET framework and in section~\ref{sec:calc-details} we present details of
our calculation. Section~\ref{sec:results} contains four-loop results for the
calculated heavy quark field anomalous dimension as well as small angle
expansion of QCD cusp anomalous dimension. We conclude in
section~\ref{sec:conclusion}.

\section{Cusp anomalous dimension in HQET framework}
\label{sec:HQET-setup}

Heavy Quark Effective Theory is a well established framework for calculation of
both full angle dependent
$\cusp(\phi,\alpha_s)$~\cite{Grozin:2014hna,Grozin:2015kna,Bruser:2020bsh} and
its small-angle expansion~\cite{Grozin:2017css,Bruser:2019auj}. In our work we
closely follow the technique employed in
Refs.~\cite{Grozin:2017css,Bruser:2019auj} for the calculation of
$\cusp(\phi,\alpha_s)$ in small angle expansion. Within the HQET framework the
quantity $\cusp(\phi,\alpha_s)$ is extracted from UV divergences of diagrams
with cusped HQET line where the cusp corresponds to an abrupt change of heavy
quark velocity.
Being an effective theory of QCD, HQET describes the interaction of heavy quark
field $h$ with massless quarks by gluon exchange. The renormalization of HQET
theory requires, in addition to QCD renormalization constants, the only new
constant $Z_h$ connecting the bare and the renormalized heavy quark fields,
which appears in the diagrams with external heavy legs. The QCD renormalization
constants up to the four-loop order can be found in Ref.~\cite{Czakon:2004bu}.
To fix notation we provide Feynman rules for the heavy quark propagator and the
heavy-quark-gluon vertex:
\begin{align}
  \label{eq:HQET-feyn-rules}
  & \vcenter{\hbox{
    \begin{tikzpicture}
      \coordinate (a) at (-1,0); \coordinate (b) at (1,0);
      \draw[WLB] (a)  --  node[pos=0.15, anchor = north] {$v$} (b)  {};
      \draw[-{Stealth[scale=1]}] (-0.5,0.2)  --  node[pos=0.5, anchor = south] {$p$} (0.5,0.2)  {};
      \node[anchor=south] at (a) {\small$i$};
      \node[anchor=south] at (b) {\small$j$};
    \end{tikzpicture}
    }}
    = \frac{-i \delta_{i j}}{\omega - v \cdot p} \; ,
  &
  &
    \vcenter{\hbox{
    \begin{tikzpicture}
      \coordinate (a) at (-1,0); \coordinate (b) at (1,0); \coordinate (c) at (0,0.55);
      \draw[WLB] (a)  --  node[pos=0.15, anchor = north] {$v$} (b)  {};
      \draw[GLUON] (0,0.03) -- (c);
      \fill (0,0) circle (1pt);
      \node[anchor=south] at (a) {\small$i$};
      \node[anchor=south] at (b) {\small$j$};
      \node[anchor=west] at (c) {\small$\mu,a$};
    \end{tikzpicture}
    }}
    = ig v^\mu T_{ij}^a \; . &
\end{align}
By introducing in Eq. \eqref{eq:HQET-feyn-rules} the residual energy $\omega$ we
regulate the IR divergencies in the diagrams, then the UV divergences, our main
interest, reveal themselves as poles in $\ep=(4-d)/2$. Since one of our goals,
$h$-field anomalous dimension, is known to be gauge dependent quantity we
perform our calculations in the $R_\xi$-gauge where the gluon propagator has the
form
\begin{equation}
  \label{eq:gluon-prop}
  \vcenter{\hbox{
      \begin{tikzpicture}
        \coordinate (a) at (-1,0); \coordinate (b) at (1,0);
        \draw[GLUON] (a) -- (b);
        \draw[-{Stealth[scale=1]}] (-0.5,0.2)  --  node[pos=0.5, anchor = south] {$p$} (0.5,0.2)  {};
        \node[anchor=south] at (a) {\small$a$};
        \node[anchor=south] at (b) {\small$b$};
      \end{tikzpicture}
    }}
  =\frac{-i\delta_{ab}}{p^2}\left[ g_{\mu\nu} - \xi \frac{p_\mu p_\nu}{p^2}\right]\,.
\end{equation}
All other QCD Feynman rules are standard and available from Ref.~\cite{Grozin:2007zz}.
Equipped with the above Feynman rules we are prepared to calculate loop diagrams
in HQET. We perform our calculations in two steps. In the first step, we
calculate the sum of bare unrenormalized diagrams up to the needed order. In the
second step, we carry out renormalization and extract the required
renormalization constants.
The first and simplest quantity we need to calculate is the heavy quark
field renormalization constant and the corresponding anomalous dimension
$\gamma_h$. We consider the quantity
\begin{equation}
  \label{eq:Ghh-def}
  \Ghh  = \vcenter{\hbox{
      \begin{tikzpicture}
        \draw[WLBS] (-1.5,-1.5) -- node[pos=0.12, anchor = north] {$v$} (1.5,-1.5)  {};
        \draw[color=colD,opacity=0.5] (1,-1.45) arc (0:180:1);
        \fill[pattern=north west lines, pattern color=colD,opacity=0.2] (1,-1.45) arc (0:180:1);
        \node[anchor=east] at (-1.5,-1.5) {$i$};
        \node[anchor=west] at (1.5,-1.5) {$j$};
        \path[use as bounding box] (-2,-2.5) rectangle (2,0.0);
      \end{tikzpicture}
    }} \cdot \frac{\delta_{ij}}{\omega N}.
\end{equation}
where $N$ is the number of colors and the dashed blob stands for the sum of the bare 1PI two-point functions up to the four-loop order. For convenience, we contract the color indices to make
the expression scalar and choose the normalization factor such that the
perturbative expansion of $\Ghh$ starts with 1.
Another quantity that we want to calculate is the four-loop QCD beta-function.
Although this beta-function is known for a long
time~\cite{vanRitbergen:1997va,Czakon:2004bu}, this calculation provides a
crucial check of our setup. We extract this quantity from the heavy-quark-gluon
vertex. Similar to the previous case, Eq. \eqref{eq:Ghh-def}, we define the
scalar function $\Gghh$ accumulating the contributions of 1PI three-point
diagrams up to four-loop order:
\begin{equation}
  \label{eq:Gghh-def}
  \Gghh  = \vcenter{\hbox{
      \begin{tikzpicture}
        \draw[WLBS] (-1.5,-1.5) -- node[pos=0.12, anchor = north] {$v$} (1.5,-1.5)  {};
        \draw[GLUON] (0,-0.45) -- (0.0,0.09);
        \fill[pattern=north west lines, pattern color=colD,opacity=0.2] (-1,-1.45)--(0,-0.45)--(1,-1.45)--cycle;
        \draw[color=colD,opacity=0.5] (-1,-1.45) -- (0,-0.45) -- (1,-1.45);
        \node[anchor=east] at (-1.5,-1.5) {$i$};
        \node[anchor=west] at (1.5,-1.5) {$j$};
        \node[anchor=west] at (0.3,0) {$a,\mu$};
        \fill (0,-0.45) circle (1pt);
        \path[use as bounding box] (-2,-2.5) rectangle (2,0.8);
      \end{tikzpicture}
    }} \cdot \frac{t^a_{ij} v_{\mu}}{g_s N C_F}.
\end{equation}
To reduce the original problem to the problem of propagator-type diagrams
calculation, we apply the IRR trick~\cite{Vladimirov:1979zm} to diagrams
entering in~\eqref{eq:Gghh-def} and put external gluon momentum to zero.
Finally, we want to consider the cusp anomalous dimension and, in particular,
calculate its small angle expansion. We consider the expectation value of an
infinite cusped Wilson line depending on two velocities $v^2=v^{\prime 2}=1, v
\cdot v^{\prime} = \cos{\phi}$. In momentum representation the perturbative
corrections to this expectation value are expressed via HQET diagrams with
$[h(v)\bar{h}(v')]$ operator insertion into two-point function with heavy quark
velocities $v$ and $v^{\prime}$ to the left and to the right from the insertion
point, respectively. Again, we construct a scalar function $\Ghhhh(\phi)$
corresponding to the sum of bare 1PI diagrams convoluted with the appropriate
tensor:
\begin{equation}
  \label{eq:Gcusp-def}
  \Ghhhh(\phi) = \vcenter{\hbox{
      \begin{tikzpicture}
        \draw[WLBE] (-1.5,-1.5)--(0,0) -- (1.5,-1.5);
        \draw[dashed] (0,0) -- (0.5,0.5);
        \draw (0.3,-0.28) arc (-30:35:0.5);
        \fill[pattern=north west lines, pattern color=colD,opacity=0.2] (-1,-1.05)--(0,-0.05)--(1,-1.05)--cycle;
        \draw[color=colD,opacity=0.5] (-1,-1.05) -- (1,-1.05);
        \node[anchor=east] at (-1.5,-1.5) {\small$i$};
        \node[anchor=north west] at (-1.5,-1.5) {$v$};
        \node[anchor=west] at (1.5,-1.5) {\small$j$};
        \node[anchor=north east] at (1.5,-1.5) {$v^{\prime}$};
        \node[anchor=west] at (0.3,0) {$\phi$};
        \path[use as bounding box] (-2,-2.5) rectangle (2,0.8);
      \end{tikzpicture}
    }} \cdot \frac{\delta_{ij}}{N}.
\end{equation}
Using results for bare functions $G_X$ calculated before we proceed with the
second step, namely, with the extraction of renormalization constants. The
corresponding $\MS$ renormalization constants $Z_X$ for each function $G_X$ are
determined from the poles cancellation requirements
\begin{align}
  \label{eq:Zvert-def}
  &
  \Zhh \cdot \Ghh = \mathcal{O}\left( \ep^0 \right),
  & &
  \Zghh \cdot \Gghh = \mathcal{O}\left( \ep^0 \right),
  & &
  \Zhhhh (\phi) \cdot \Ghhhh(\phi) = \mathcal{O}\left( \ep^0 \right).
\end{align}
Here we replace bare parameters $a_{s,B}$ and $a_{\xi,B}=1-\xi$ entering
$G_i$ with its renormalized counterparts:
\begin{align}
  \label{eq:Za-def}
  & a_{s,B} = \mu^{2\ep} Z_{\aS} \aS,  & &   a_{\xi,B} = Z_{\aXi} \aXi.
\end{align}
From vertex renormalization constants~\eqref{eq:Zvert-def}, dividing by external
legs $Z$-factors we determine gauge parameter independent combinations:
\begin{align}
  \label{eq:ZcuspZas-def}
  &
    Z_{\aS} = \frac{\Zghh^2}{\Zhh^2 Z_A},
  & &
      \Zcusp(\phi) = \frac{\Zhhhh(\phi)}{\Zhh},
\end{align}
where $\Zhh$ found before and gluon field renormalization constant $Z_A$ is
known from Refs.~\cite{vanRitbergen:1997va,Czakon:2004bu}. Gauge parameter
independence of $Z_{\aS}$ and $\Zcusp(\phi)$ allows us to calculate $\Gghh$ and
$\Ghhhh(\phi)$ as expansion around $\xi=0$, keeping only the first term of
expansion to verify its cancellation in \eqref{eq:ZcuspZas-def} as additional
test on the validity of the obtained results. Another test comes from the HQET
Ward identity implying that:
\begin{equation}
  \label{eq:Zphi0-eq} \Zhh = \lim_{\phi\to0}{\Zhhhh}(\phi).
\end{equation}
From the four-loop result for $Z_{\aS}$ in \eqref{eq:ZcuspZas-def} we can derive
a well known expression for the four-loop QCD beta-function within the $\MS$
renormalization scheme:
\begin{equation}
  \label{eq:betaQCD-def}
  \beta_{\aS} = \frac{d \aS}{d \log{\mu^2}} = \frac{-\ep \aS }{1+ \aS \partial_{\aS}\log{Z_{\aS}}} = -\ep \aS - \sum\limits_{n=0}^{\infty} b_n \aS^{n+2},
\end{equation}
with $b_0 = \frac{11}{3} C_A - \frac{4}{3} n_f T_F$. The agreement of the
obtained coefficients $b_{0-3}$ with the results of
Refs.~\cite{vanRitbergen:1997va,Czakon:2004bu} provides a strong check of our
calculation setup.

\section{Calculation details}
\label{sec:calc-details}
To calculate bare Green functions introduced in section~\ref{sec:HQET-setup} we
have developed a highly automatized setup. Its workflow starts with the
generation of diagrams with \texttt{DIANA}~\cite{Tentyukov:1999is}, which
internally calls \texttt{QGRAF}~\cite{Nogueira:1991ex}. We generate the
propagator-type (two-point) diagrams for the calculation of $\Ghh$ and
vertex-type (three-point) diagrams for the calculation of $\Gghh$. The former
diagrams have been reused in $\Ghhhh$ calculation since the diagrams with the
cusp on the Wilson line are in one-to-one correspondence with the diagrams
obtained by an auxiliary leg insertion in all possible ways on the heavy-quark
line in two-point diagrams. The insertion point corresponds to a cusp, so we
replace $v\to v^{\prime}$ in all $h$-propagators to the right of this point.
After that, all propagators dependent on $v^{\prime}$ are expanded in the
vicinity of $\phi=0$ with a recursive application of the identity
\begin{equation}
  \label{eq:hqProp-exp}
  \frac{1}{1-2\, k \cdot v^{\prime}} =   \underbrace{\frac{1}{1-2\, k \cdot v}}_{\mathcal{O}(\phi^0)}
  + \underbrace{\frac{1}{1-2\, k \cdot v} \frac{2 k \cdot (v^{\prime} - v)}{1-2\, k \cdot v^{\prime}}}_{\mathcal{O}(\phi)}
\end{equation}
After the decomposition of $v^{\prime}$ in numerator with $v^{\prime} = v
\cos{\phi} + n_{\perp} \sin{\phi}$, where $n_{\perp}^2 = 1$, and $v \cdot
n_{\perp} = 0$ we are left with scalar products of $n_{\perp}$ with loop
momenta. Since the result of the loop integration is independent of the
$n_{\perp}$ direction it is possible to replace $n_{\perp}^{\mu_1} \dots
n_{\perp}^{\mu_{n}}\to \left\langle n_{\perp}^{\mu_1} \dots
n_{\perp}^{\mu_{n}}\right\rangle$, where $\langle\bullet\rangle$ denotes
averaging over perpendicular directions. For reference, we present explicit
formulae for this averaging
\begin{equation}
  \label{eq:vT-aver}
  \left\langle n_{\perp}^{\mu_1} \dots n_{\perp}^{\mu_{2s-1}}\right\rangle = 0,\quad
  \left\langle n_{\perp}^{\mu_1} \dots n_{\perp}^{\mu_{2s}}\right\rangle= \frac{(1/2)_s}{(3/2 - \ep )_s}\mathcal{S} \prod_{k=1}^{s} g_\perp^{\mu_{2k-1}\mu_{2k}}\,,
\end{equation}
where $g_\perp^{\alpha\beta}=g^{\alpha\beta}-v^\alpha v^\beta$, $c_s = c\cdot
(c+1)\cdot\ldots \cdot(c+s-1)$ is the Pocchammer symbol, and $\mathcal{S}$ is
the normalized (i.e., $\mathcal S1=1$) symmetrization operator with respect to
permutations of $\mu_1,\ldots,\mu_{2k}$. From Eq. \eqref{eq:hqProp-exp} it is
obvious that the calculation of higher orders of expansion in $\phi$ requires
the reduction of integrals with higher powers of denominators and scalar
products in the numerator. In our work we consider expansion to $\phi^4$,
corresponding to two first non-trivial orders in the small-angle expansion of
$\cusp(\phi)$.
Next, we calculate the Dirac traces and simplify expressions with
\texttt{FORM}~\cite{Kuipers:2012rf} and perform the color algebra in terms of
color invariants with \texttt{COLOR}~\cite{vanRitbergen:1998pn} ending up with a
set of scalar integrals.
Due to the presence of linear propagators~\eqref{eq:HQET-feyn-rules}, we need to
perform partial fraction decomposition of linear dependent propagators. With the
implementation based on the package \texttt{TopoID}~\cite{Hoff:2016pot} and the
private version of the \texttt{LiteRed} package, we obtain expressions containing
integrals with an independent set of scalar products only which can be mapped on
the set of 19 auxiliary topologies considered in~\cite{Lee:2022}. For reduction
to master integrals calculated in~\cite{Lee:2022}, we use
\texttt{FIRE6}~\cite{Smirnov:2019qkx} in combination with
\texttt{LiteRed}~\cite{Lee:2012cn,Lee:2013mka}.
All diagrams up to the three-loop order as well as the four-loop diagrams needed
for $\Ghh$ are calculated keeping the full dependence on gauge-fixing parameter
$\xi$, but to reduce required calculation time, we perform expansion in $\xi$ to
leading order in four-loop diagrams $\Gghh$ and $\Ghhhh$, since corresponding
renormalization constants~\eqref{eq:ZcuspZas-def} extracted from these functions
are gauge-parameter independent and cancelation of the $\xi$ dependence in the
expanded form is a sufficient check on the validity of the obtained result.

\section{Results and discussion}
\label{sec:results}
From the results for renormalization constants obtained in the previous section,
we derive anomalous dimensions by taking logarithmic derivatives in the
renormalization scale\footnote{For historical reasons, we take derivatives in
  $\log{\mu}$ rather than in $\log{\mu^2}$ for $\gamma_h$ and $\cusp(\phi)$.}:
\begin{align}
  \gamma_h  & = \frac{d \log \Zhh}{d \log{\mu}} =
              2 \beta_{\aS}\frac{\partial \log{\Zhh}}{\partial \aS} + 2 \beta_{\aXi}\frac{\partial \log{\Zhh}}{\partial \aXi},    \label{eq:gamH-def}
  \\
  \cusp(\phi) & = -\frac{d \log Z_{\rm cusp}(\phi)}{d \log{\mu}} = - 2 \beta_{\aS} \frac{d \log{Z_{\rm cusp}(\phi)}}{d \aS}.
                \label{eq:gamCusp-def}
\end{align}
Here $a_s=\alpha_s/(4\pi)$ and $a_{\xi}=1-\xi$, the strong coupling
beta-function $\beta_{\aS}$ was introduced in~\eqref{eq:betaQCD-def} and the
beta-function of the gauge fixing parameter $\aXi$ is defined as follows:
\begin{equation}
  \label{eq:betaXI-def}
  \beta_{\aXi} = \frac{d \aXi}{d \log{\mu^2}} = \beta_{\aS}\frac{-\aXi \partial_{\aXi} \log{Z_{\aXi}} }{1+ \aXi \partial_{\aXi} \log{Z_{\aXi}}}\,.
\end{equation}
The complete result for the HQET field anomalous dimension up to four-loop order is
\allowdisplaybreaks
\begin{align}
  \label{eq:gamH-res}
  \gamma_h =
  & -2 \aS \CF (3 - \aXi)
    + \aS^2 \CF\left\{
    \frac{32}{3} \TFNF  - \CA \left( \frac{179}{6} - 4 \aXi - \frac{1}{2}\aXi^2 \right)  \right\}
    \nonumber\\
  & +\aS^3\left\{
    \CFCF \TFNF \left(102 - 96 \zeta_3\right) + \CF \left( \frac{160}{27} (\TFNF)^2 +
    \CA \TFNF \left( \frac{782}{27} + 96 \zeta_3 - \frac{17}{2} \aXi \right)
    \right.\right.\nonumber\\
  & -\left.\left.
    \CACA \left(
    \frac{23815}{216}
    + \frac{123}{4}\zeta_3
    + \frac{4}{15} \pi^4
    - \left(\frac{271}{16} - \frac{4}{45} \pi^4 + 6 \zeta_3\right) \aXi
    -\left(\frac{39}{16} + \frac{3}{4}\zeta_3\right)\aXi^2
    - \frac{5}{8}\aXi^3
    \right)
    \right)
    \right\}
    \nonumber\\
  & + \aS^4 \left\{
    \CFCF \left[
    (\TFNF)^2 \left(-\frac{3296}{27} - \frac{32}{15}\pi^4 + 384 \zeta_3\right)
    \right.\right.
    \nonumber\\
  & \left. + \CA \TFNF
    \left(\frac{21703}{27}
    + \frac{88}{15}\pi^4
    - 928 \zeta_3
    - 480 \zeta_5
    - \left(\frac{767}{6} - \frac{4}{15}\pi^4 - 88 \zeta_3 \right) \aXi
    \right)
    \right]
    \nonumber\\
  & -\frac{\dFdF}{\NC} \NF \left(
    \frac{512}{3} \pi^2
    - 256 \zeta_3
    - \frac{512}{3}\pi^2\zeta_3
    + 320 \zeta_5
    \right)
    - \CFCFCF \TFNF \left(
    \frac{560}{3}
    + 592 \zeta_3
    - 960 \zeta_5
    \right)
    \nonumber\\
  &  + \CF \left[
    (\TFNF)^3 \left(\frac{256}{27} - \frac{256}{9} \zeta_3\right)
    - \CA (\TFNF)^2  \left(
    \frac{2054}{81}
    - \frac{32}{15}\pi^4
    + 384\zeta_3
    + \left(\frac{2152}{243} - \frac{32}{3}\zeta_3\right) \aXi
    \right)
    \right.
    \nonumber\\
  & + \CACA \TFNF \left(
    \frac{30617}{81}
    - \frac{16}{3} \pi^2
    - \frac{3097}{540}\pi^4
    + \frac{5506}{3}\zeta_3
    + \frac{104}{9} \pi^2 \zeta_3 - 96 \zeta_3^2
    - \frac{1534}{3}\zeta_5\right.
    \nonumber\\
  & \left.- \left(
    \frac{37957}{1944}
    - \frac{1}{15}\pi^4
    + 82\zeta_3
    + \frac{16}{27} \pi^2 \zeta_3
    - \frac{4}{9}\zeta_5
    \right) \aXi
    - \left(
    \frac{109}{36}
    - \frac{1}{180}\pi^4
    + \frac{7}{3}\zeta_3
    \right) \aXi^2
    \right)
    \nonumber\\
  & - \CACACA \left(
    \frac{471001}{648}
    - \frac{781}{36}\pi^2
    + \frac{10501}{2160}\pi^4
    - \frac{850}{1701} \pi^6
    + \frac{212237}{288}\zeta_3
    + \frac{709}{36}\pi^2\zeta_3
    - \frac{451}{4} \zeta_3^2
    - \frac{3859}{12}\zeta_5
    \right.
    \nonumber\\
  &
    - \left(
    \frac{1690475}{15552}
    - \frac{2}{9}\pi^2
    - \frac{9109}{4320}\pi^4
    + \frac{472}{8505}\pi^6
    + \frac{3839}{48}\zeta_3
    + \frac{164}{27}\pi^2\zeta_3
    + \frac{11}{4}\zeta_3^2
    - \frac{272}{9}\zeta_5
    \right)\aXi
    \nonumber\\
  & - \left(\frac{6707}{576} + \frac{1}{12}\pi^2 - \frac{121}{1080}\pi^4
    + \frac{653}{48}\zeta_3 + \frac{13}{36} \pi^2 \zeta_3 - \frac{169}{48}\zeta_5\right) \aXi^2
    - \left(\frac{149}{48} - \frac{1}{480}\pi^4 + \frac{21}{16}\zeta_3\right) \aXi^3
    \nonumber\\
  & \left.\left. - \left(\frac{19}{32} + \frac{1}{96}\zeta_3 + \frac{5}{48}\zeta_5 \right) \aXi^4
    \right)
    \right]
    \nonumber\\
  & - \frac{\dFdA}{\NC} \left[
    \frac{16}{3}\pi^2
    - \frac{128}{15}\pi^4
    - \frac{224}{405}\pi^6
    - \frac{569}{4}\zeta_3
    - \frac{320}{3}\pi^2\zeta_3
    + 384 \zeta_3^2
    + \frac{4815}{4}\zeta_5\right.
    \nonumber\\
  & - \left(\frac{16}{3}\pi^2 - \frac{884}{2835}\pi^6 - 11 \zeta_3 - \frac{128}{3}\pi^2\zeta_3
    + 24 \zeta_3^2 + 540 \zeta_5\right)\aXi
    - \left(\frac{3}{2}\zeta_3 - \frac{75}{2}\zeta_5 \right)\aXi^2
    \nonumber\\
  & \left.\left.+ 3 \zeta_3 \aXi^3 
    + \left(\frac{7}{4} \zeta_3 - \frac{5}{4}\zeta_5\right) \aXi^4
    \right]
    \right\}
    + \mathcal{O}\left( \aS^5 \right)
\end{align}
The terms up to $\alpha_s^3$ agree with the results of the three-loop HQET
calculation~\cite{Chetyrkin:2003vi} and results obtained as a byproduct of the
three-loop QCD on-shell renormalization~\cite{Melnikov:2000zc}. The analytical
result for the four-loop part is new and in full agreement with the results of
numerical calculation~\cite{Marquard:2018rwx} and partial four-loop results for
fermionic contributions calculated in~\cite{Grozin:2017css,Bruser:2019auj}. We
note also that the terms proportional to $\aXi^{L-k}$ with $L\geqslant 2$ and
$k=0,1$ coincide, including the new terms for $L=4$, with the corresponding
terms in quark anomalous dimension $\gamma_q$, \cite{Chetyrkin2017}.
The main result of the present paper, namely the small angle expansion of the
cusp anomalous dimension has the following form:
\begin{equation}
  \label{eq:gam-cusp-exp}
  \cusp(\phi) = \Gamma^{(2)} \phi^2 + \Gamma^{(4)} \phi^4 + \mathcal{O}(\phi^6)
\end{equation}
and the complete results for the first two terms of the small-angle expansion of
the cusp anomalous dimension up to four-loop order derived
from~\eqref{eq:gamCusp-def} read
\begin{align}
  \label{eq:gamCuspPhi2-res}
  \Gamma^{(2)}
  & =
    - \frac{4}{3} \aS \CF
    - \aS^2 \CF \left\{  \CA
    \left(
    \frac{376}{27}
    - \frac{8}{9}\pi^2
    \right)
    - \frac{80}{27} \TFNF
    \right\} \nonumber\\
  & + \aS^3\left\{  \CFCF \TFNF \left(
    \frac{220}{9}
    - \frac{64}{3}\zeta_3
    \right)
    + \frac{64}{81}\CF (\TFNF)^2 \right.
    \nonumber\\
  & \qquad - \CF \CACA \left(
    \frac{946}{9}
    - \frac{1360}{81}\pi^2
    + \frac{8}{9}\pi^4
    + \frac{40}{9}\zeta_3
    \right)
    \nonumber\\
  & \qquad \left.+ \CF \CA \TFNF \left(
    \frac{3112}{81}
    - \frac{320}{81}\pi^2
    + \frac{224}{9}\zeta_3
    \right)
    \right\}
    \nonumber\\
  & + \aS^4 \left\{
    \frac{\dFdF}{\NC} \NF \left(
    \frac{640}{27} \pi^2
    + \frac{320}{27} \pi^4
    - \frac{1024}{9}\pi^2 \zeta_3
    \right)\right.
    \nonumber\\
  & \qquad + \frac{\dFdA}{\NC} \left(
    \frac{64}{27}\pi^2
    - \frac{512}{27}\pi^4
    - \frac{128}{135}\pi^6
    + \frac{2176}{9}\pi^2\zeta_3
    \right)
    \nonumber\\
  & \qquad 
    - \CFCF (\TFNF)^2 \left(
    \frac{9568}{243}
    + \frac{64}{135}\pi^4
    - \frac{2560}{27}\zeta_3
    \right)
    + \CF (\TFNF)^3\left(
    \frac{256}{243}
    - \frac{512}{81}\zeta_3
    \right)
    \nonumber\\
  & \qquad + \CFCF \CA \TFNF \left(
    \frac{103772}{243}
    - \frac{880}{27}\pi^2
    + \frac{176}{135}\pi^4
    -\frac{10880}{27} \zeta_3
    + \frac{256}{9}\pi^2\zeta_3
    - \frac{320}{3}\zeta_5
    \right)
    \nonumber\\
  &  \qquad
    -\CF \CA (\TFNF)^2\left(
    \frac{7340}{243}
    - \frac{2432}{729}\pi^2
    - \frac{224}{405}\pi^4
    +\frac{8960}{81}\zeta_3
    \right)
    \nonumber\\
  & \qquad + \CF \CACA \TFNF\left(
    \frac{96322}{243}
    - \frac{59072}{729}\pi^2
    + \frac{352}{81}\pi^4
    + \frac{57776}{81}\zeta_3
    - \frac{896}{27}\pi^2\zeta_3
    - \frac{1760}{9}\zeta_5
    \right)
    \nonumber\\
  & \qquad - \CF \CACACA\left(
    \frac{178022}{243}
    - \frac{143624}{729}\pi^2
    + \frac{9682}{405}\pi^4
    - \frac{80}{81}\pi^6
    + \frac{19024}{81}\zeta_3
    - \frac{256}{27}\pi^2\zeta_3
    - \frac{1240}{9}\zeta_5\right)
    \nonumber\\
  & \qquad \left. - \CFCFCF\TFNF\left(
    \frac{1144}{27}
    + \frac{1184}{9}\zeta_3
    - \frac{640}{3}\zeta_5
    \right)
    \right\}
    +\mathcal{O}\left( \aS^5 \right),
\end{align}
\begin{align}
  \label{eq:gamCuspPhi4-res}
  \Gamma^{(4)}
  & =
    -\frac{4}{45} \aS \CF
    - \aS^2 \CF \left\{  \CA
    \left(
    \frac{364}{405}
    - \frac{8}{135}\pi^2
    \right)
    - \frac{16}{81}\TFNF\right\}
    \nonumber\\
  & + \aS^3\left\{ \CFCF\TFNF\left(
    \frac{44}{27}
    - \frac{64}{45}\zeta_3
    \right)
    + \frac{64}{1215}\CF(\TFNF)^2\right.
    \nonumber\\
  & \qquad - \CF \CACA \left(
    \frac{18074}{6075}
    - \frac{320}{243}\pi^2
    + \frac{8}{135}\pi^4
    + \frac{3512}{675}\zeta_3
    \right)
    \nonumber\\
  & \qquad \left.+ \CF \CA\TFNF \left(
    \frac{328}{135}
    - \frac{64}{243}\pi^2
    + \frac{224}{135}\zeta_3
    \right)
    \right\}  
    \nonumber\\
  & +  \aS^4 \left\{
    \frac{\dFdF}{\NC}\NF \left(
    -\frac{1472}{225}
    - \frac{10048}{2025}\pi^2
    + \frac{3136}{2025}\pi^4
    + \frac{18176}{225}\zeta_3
    - \frac{4096}{675}\pi^2 \zeta_3
    - \frac{1024}{9}\zeta_5
    \right)\right.
    \nonumber\\
  & \qquad + \frac{\dFdA}{\NC} \left(
    \frac{512}{243}
    - \frac{53696}{3645}\pi^2
    - \frac{1984}{675}\pi^4
    - \frac{128}{2025}\pi^6
    +\frac{36224}{405}\zeta_3
    + \frac{9344}{225}\pi^2\zeta_3
    - \frac{896}{9}\zeta_5
    \right)
    \nonumber\\
  & \qquad -\CFCF(\TFNF)^2 \left(
    \frac{9568}{3645}
    + \frac{64}{2025}\pi^4
    - \frac{512}{81}\zeta_3
    \right)
    + \CF (\TFNF)^3\left(
    \frac{256}{3645}
    - \frac{512}{1215}\zeta_3
    \right)
    \nonumber\\
  & \qquad + \CFCF \CA \TFNF\left(
    \frac{99812}{3645}
    - \frac{176}{81}\pi^2
    + \frac{176}{2025}\pi^4
    - \frac{10496}{405}\zeta_3
    + \frac{256}{135}\pi^2\zeta_3
    - \frac{64}{9}\zeta_5
    \right)
    \nonumber\\
  & \qquad 
    - \CF \CA(\TFNF)^2 \left(
    \frac{20804}{10935}
    - \frac{2432}{10935}\pi^2
    - \frac{224}{6075}\pi^4
    + \frac{1792}{243}\zeta_3
    \right)
    \nonumber\\
  & \qquad + \CF \CACA\TFNF \left(
    \frac{224414}{273375}
    - \frac{286696}{54675}\pi^2
    + \frac{4688}{30375}\pi^4
    + \frac{2389232}{30375}\zeta_3
    - \frac{1664}{675} \pi^2\zeta_3
    - \frac{1504}{135}\zeta_5
    \right)
    \nonumber\\
  & \qquad + \CF \CACACA\left(
    \frac{9434794}{273375}
    + \frac{216896}{18225}\pi^2
    - \frac{5782}{3375}\pi^4
    + \frac{16}{243}\pi^6
    - \frac{5252768}{30375}\zeta_3
    + \frac{2912}{675}\pi^2\zeta_3
    + \frac{4168}{45}\zeta_5
    \right)
    \nonumber\\
  & \left.\qquad
    - \CFCFCF\TFNF \left(
    \frac{1144}{405}
    + \frac{1184}{135} \zeta_3
    - \frac{128}{9}\zeta_5
    \right)
    \right\}
    +\mathcal{O}\left( \aS^5 \right)
\end{align}

The terms up to $\alpha_s^3$ agree with the full angle-dependent
results~\cite{Grozin:2014hna,Grozin:2015kna} expanded in $\phi^2$. Four-loop
part is new and its fermionic contributions are in agreement with the partial
four-loop results from~\cite{Grozin:2017css,Bruser:2019auj}. Results
\eqref{eq:gamCuspPhi2-res} and \eqref{eq:gamCuspPhi4-res} are obtained for the
QCD-like case, where heavy quark and massless quarks are in the same
representation. For the case of different representation $R$ of the Wilson line,
the result can be easily modified by exchanging the single power of $C_F$ with
$C_R$ and by replacements $\frac{d^{abcd}_Fd^{abcd}_F}{N} \to
\frac{d^{abcd}_Rd^{abcd}_F}{N_R}$ and $\frac{d^{abcd}_Fd^{abcd}_A}{N} \to
\frac{d^{abcd}_Rd^{abcd}_A}{N_R}$ for quartic Casimir invariants.
It is interesting to compare our result for QCD $\cusp$ small-angle expansion
with available results in $\mathcal{N}=4$ SYM. In particular, we compare the
Bremsstrahlung function which is known in $\mathcal{N}=4$ SYM as an all-order
expression \cite{Correa:2012at}. By retaining in front of $\aS^L$ only the terms
of highest transcendental weight $2L-2$ in QCD Bremsstrahlung function $B^{\rm
QCD}=-\Gamma^{(2)}$ from Eq. \eqref{eq:gamCuspPhi2-res} we obtain
\begin{align}
  \label{eq:B-max-tr}
  &   B_{\rm MT}^{\rm QCD} =
    \frac{4}{3} \CF \aS
    - \frac{8}{9}  \CF \CA \pi^2 \aS^2
    + \frac{8}{9} \CF \CACA \pi^4 \aS^3
    - \left\{\frac{80}{81}  \CF \CACACA   - \frac{128}{135}  \frac{\dFdA}{\NC}  \right\} \pi^6 \aS^4\,.
\end{align}
Note that all color factors entering $B_{\rm MT}^{\rm QCD}$ are of the maximal
non-abelian nature.
All-order expression for the $\BSYM$ from Ref. \cite{Correa:2012at}
reads\footnote{Note that scalar field contribution depends on auxiliary angle
$\theta$ in ``inner'' space, and Eq. \eqref{eq:B-N4SYM} corresponds to the case
$\theta=0$.}
\begin{equation}
  \label{eq:B-N4SYM}
  \BSYM = \frac{\aS}{2 \pi^2} \partial_{\aS} \log {\left[  L_{N - 1}^{(1)}\left( -4 \pi^2 \aS\right) e^{2 \pi^2 \aS (1 - 1/N)}\right]},
\end{equation}
where $L_n^{(\alpha)}$ is the generalized Laguerre polynomial. This
Bremsstrahlung function, in addition to the contribution of gluons, also
involves the contribution of auxiliary scalar fields which leads to different
results for $B^{\rm QCD}$ and $\BSYM$ already at one loop: they differ by a
factor of $3/2$. Remarkably, the same relation holds at least to the four
loops when we replace $B^{\rm QCD}$ by its maximal transcendentality part:
\begin{equation}
  \label{eq:Bratio}
  \BSYM = \frac{3}{2} B_{\rm MT}^{\rm QCD} + \mathcal{O}\left(\aS^5\right)\,.
\end{equation}
Moreover, we have checked that the above relation also holds for arbitrary
representation of Wilson line once we substitute $C_F\to C_R$ and
$d_F^{abcd}d_A^{abcd}\to d_R^{abcd}d_A^{abcd}$ in Eq. \eqref{eq:B-max-tr} and
use the perturbative result of Ref. \cite{Fiol:2018yuc} for $\BSYM$ in
representation $R$. The relation \eqref{eq:Bratio} can be interpreted as the
manifestation of the maximal transcendentality
principle~\cite{Kotikov:2002ab,Kotikov:2004er}.
Finally, let us write the small-angle expansion of $\cusp$ in a slightly modified form
\begin{equation*}
  \cusp(\phi)=-3{\Gamma}^{(2)}A(x)+\widetilde{\Gamma}^{(4)}\phi^4+\mathcal{O}\left(\phi^6\right) =
  {\Gamma}^{(2)}\cdot \left(\phi^2+\frac{\phi^4}{15}\right)+\widetilde{\Gamma}^{(4)}\phi^4+\mathcal{O}\left(\phi^6\right),
\end{equation*}
where $x=e^{i\phi}$ and $A(x)=\phi\cot\phi-1$ is simply the angular dependence
of one-loop cusp anomalous dimension in QCD. The modified coefficient
$\widetilde{\Gamma}^{(4)}={\Gamma}^{(4)}-\frac1{15}{\Gamma}^{(2)}$ has a
substantially simpler form than ${\Gamma}^{(4)}$ in Eq.
\eqref{eq:gamCuspPhi4-res}:
\begin{align}
  \widetilde{\Gamma}^{(4)}
  &=
    \frac{4}{135}\CA \CF\aS^2
    + \aS^3 \CF \left\{
    \CACA\left(\frac{24496}{6075} + \frac{16}{81}\pi^2 - \frac{368}{75} \zeta_3\right)
    - \frac{32}{243} \CA\TFNF
    \right\} \nonumber\\
  & + \aS^4\left\{
    \CFCF \CA \TFNF \left(-\frac{88}{81} + \frac{128}{135}\zeta_3\right)
    + \CF\CA(\TFNF)^2\frac{1216}{10935} \right.\nonumber\\
  & + \CF\CACA\TFNF \left(-\frac{6999736}{273375} + \frac{2888}{18225}\pi^2 - \frac{4112}{30375}\pi^4
    + \frac{314944}{10125}\zeta_3 - \frac{512}{2025}\pi^2\zeta_3 + \frac{256}{135}\zeta_5\right) \nonumber\\
  & + \CF\CACACA\left(\frac{22786444}{273375} - \frac{67432}{54675} \pi^2 - \frac{3628}{30375} \pi^4
    - \frac{4777168}{30375}\zeta_3 + \frac{7456}{2025} \pi^2\zeta_3 + \frac{11264}{135}\zeta_5\right) \nonumber\\
  & + \frac{\dFdF}{\NC}\NF \left(-\frac{1472}{225} - \frac{1472}{225}\pi^2 + \frac{512}{675}\pi^4
    + \frac{18176}{225}\zeta_3 + \frac{1024}{675}\pi^2\zeta_3 - \frac{1024}{9}\zeta_5\right) \nonumber\\
  & + \left.\frac{\dFdA}{\NC}\left(\frac{512}{243} - \frac{54272}{3645} \pi^2 - \frac{3392}{2025} \pi^4
    + \frac{36224}{405}\zeta_3 + \frac{17152}{675}\pi^2\zeta_3 - \frac{896}{9}\zeta_5\right)
    \right\}
\end{align}
First, we see that $\widetilde{\Gamma}^{(4)}$ contains only one color factor
$\frac{d^{abcd}_Fd^{abcd}_F}{N}$ which survives in the QED limit. This fact is
quite anticipated agreeing with the results of
Refs.~\cite{Grozin:2018vdn,Bruser:2020bsh}, where $\cusp^{\mathrm QED}(\phi)$
was represented as $\cusp^{\mathrm QED}(\phi)=\gamma(\alpha)
A(x)+(\alpha/\pi)^4n_f B(x)$. Perhaps a less anticipated observation is that the
coefficients of all remaining color structures in $\widetilde{\Gamma}^{(4)}$ are
now free from the highest transcendental weight contribution.

\section{Conclusion}
\label{sec:conclusion}
In the present paper, we have calculated the small angle expansion of the QCD
cusp anomalous dimension and the four-loop anomalous dimension of the heavy
quark field in HQET. The obtained results agree with partial analytical and
numerical results available in the literature. We have also performed a
stringent test of our calculational setup by independent four-loop QCD
beta-function derivation from the HQET vertex renormalization. The obtained results
are the first application of the HQET propagator-type master integrals
calculated in Ref. \cite{Lee:2022}. The highly automated setup developed in the
course of this work allows one to obtain yet higher terms in the small-angle
expansion of $\cusp$ (once they are needed) as well as to calculate similar
quantities in the HQET framework. The obtained missing parts of the full QCD result
for $\cusp$ allowed us to compare the QCD result with $\mathcal{N}=4$ SYM
predictions for the Bremsstrahlung function and observe the applicability of the
maximal transcendentality principle up to four loops.
\acknowledgments
The work has been supported by Russian Science Foundation under grant
20-12-00205. We are grateful to the Joint Institute for Nuclear Research for
using their supercomputer ``Govorun.''
\bibliographystyle{JHEP}
\bibliography{cusp4l}
\end{document}